\documentclass[10pt,prl,aps,twocolumn,showpacs,floatfix]{revtex4}
\usepackage{graphicx}

\usepackage{amssymb}

\begin{document}

\title{Continuous quantum phase transition between an antiferromagnet and a valence-bond-solid in two dimensions;
evidence for logarithmic corrections to scaling}

\author{Anders W. Sandvik}
\affiliation{Department of Physics, Boston University, 590 Commonwealth Avenue, Boston, Massachusetts 02215}

\begin{abstract}
The antiferromagnetic to valence-bond-solid phase transition in the two-dimensional J-Q model (an $S=1/2$ Heisenberg model with 
four-spin interactions) is studied using large-scale quantum Monte Carlo simulations. The results support a continuous transition 
of the ground state, in agreement with the theory of ``deconfined'' quantum criticality. There are, however, large corrections to
scaling, of logarithmic or very slowly decaying power-law form, which had not been anticipated. 
This suggests that either the SU($N$) symmetric noncompact CP$^{N-1}$ field theory for deconfined 
quantum criticality has to be revised, or that the theory for $N=2$ (as in the system studied here) 
differs significantly from $N \to \infty$ (where the field theory is analytically tractable). 
\end{abstract}

\date{\today}


\maketitle

Valence-bond solid (VBS) states of two-dimensional (2D) quantum spin systems have been studied for more than two decades \cite{read} and have 
recently come into renewed focus with the theory of ``deconfined'' quantum criticality (DQC) \cite{senthil,motrunich}, which describes the 
transition between an antiferromagnetic (AF) and a VBS ground state in terms of deconfinement of spinons. In addition to the interest in such
AF--VBS transitions in condensed matter physics, there are also intriguing connections to deconfinement in gauge theories in particle physics
\cite{sachdev}. To test the validity of the DQC scenario, and to obtain quantitative results for, e.g., predicted unusual critical exponents, 
unbiased numerical studies of quantum spin hamiltonians with AF--VBS transitions are necessary.

The ``J-Q'' model was introduced recently \cite{sandvik1} as an SU($2$) symmetric spin system realizing the 2D AF--VBS transition, following earlier 
work on related U(1) symmetric models \cite{sandvik2,sandvik3}. It combines the standard Heisenberg antiferromagnet with four-spin interactions, 
which lead to local correlated bond singlets (valence bonds) and reduce the amplitudes of the longer valence bonds required \cite{liang} in 
an AF state. The J-Q model is free from ``sign problems'' \cite{signproblem}, which prohibit quantum Monte Carlo (QMC) studies of frustrated spin 
systems such as the J$_{\rm 1}$-J$_{\rm 2}$ Heisenberg model \cite{j1j2}, on which much of the past computational (exact diagonalization) research on 
VBS sates was focused. While series expansions \cite{sushkov} around various candidate states can give some insights, QMC methods \cite{evertz}, 
when applicable, are the only unbiased tools for studying 2D quantum phase transitions (in contrast to one dimension, where the 
density-matrix-renormalization-group method \cite{white} is applicable) \cite{analytical}. Being sign problem free, the J-Q model (and generalizations 
of it \cite{lou}) have opened up new avenues for exploring magnetically quantum-disordered states and quantum phase transitions. 

In this {\it Letter}, a large-scale, high-precision QMC study of the AF--VBS transition in the J-Q model is presented in order to further test the 
he DQC theory, and to settle discrepancies between previous studies \cite{sandvik1,melko,jiang}. The main point of contention is the order 
of the transition. In the DQC theory, it was argued that AF--VBS transitions are generically continuous \cite{senthil} and that the critical 
point for SU($N$) spins corresponds to a non-compact (NC) CP$^{N-1}$ field theory \cite{motrunich}. This is at odds with the long-standing 
Landau-Ginzburg paradigm, where a direct transition between two states breaking unrelated symmetries should be first-order (except at fine-tuned 
multi-critical points). Ground state \cite{sandvik1,lou} and finite-temperature \cite{melko} QMC studies of the J-Q model show scaling behavior
in good agreement with the DQC theory, including a dynamic exponent $z=1$, a rather large anomalous dimension $\eta_{\rm spin} \approx 0.35$, and 
an emergent U($1$) symmetry in the VBS phase (which in the theory is associated with spinon deconfinement). On the other hand, a QMC finite-size 
analysis by Jiang {\it et al.} would, if correct, require a first-order transition \cite{jiang}. A weakly first-order AF--VBS scenario has been 
elaborated by Kuklov {\it et al.} \cite{kuklov1,kuklov2}, based on results for a lattice model claimed to realize the NCCP$^{1}$ action, but other 
studies of the action have reached different conclusions \cite{motrunich}. 

Here it will be shown that the claimed first-order signals in the study by Jiang {\it et al.} \cite{jiang} can be attributed to over-interpretations of QMC 
data affected by significant systematical and statistical errors. The results to be presented below were obtained with the stochastic series 
expansion (SSE) method \cite{sandvik4,ssenote}, which is a finite-temperature QMC method free from systematical errors. There are no indications 
of a first-order transition, even in systems of space-time volume 20 times larger than  in \cite{jiang}. However, the data are now of high enough 
quality to detect {\it logarithmically weak deviations} from the scaling forms expected at a $z=1$ critical point. 
Logarithmic corrections are well known consequences of marginal operators at criticality, which, although they have not been predicted 
theoretically in this case (in large-$N$ treatments of the NCCP$^{N-1}$ theories \cite{senthil,nogueira,metlitski}), cannot 
{\it a priori} be ruled out for $N=2$. A first-order transition would lead to much more dramatic deviations from $z=1$.

Turning now to a quantitative discussion of the calculations, the J-Q hamiltonian \cite{sandvik1} can be written as
\begin{equation}
H = -J\sum_{\langle ij\rangle}C_{ij} - Q\sum_{\langle ijkl\rangle}C_{ij}C_{kl},
\label{ham}
\end{equation}
where $C_{ij}$ is a bond-singlet projector for $S=1/2$ spins; $C_{ij}=1/4-{\bf S}_i \cdot {\bf S}_j$. In the J term $ij$ are nearest neighbors 
on the square lattice, while $ij$ and $kl$ in the Q term are on opposite edges of a $2\times 2$ plaquette. Lattices of $N=L^2$ 
spins with periodic boundaries are used. Assuming $z=1$ (based on previous work \cite{sandvik1,melko}), the inverse temperature 
$\beta=Q/T$ is taken proportional to $L$ for finite-size scaling; $\beta=L$ and $\beta=L/4$ will be considered for $L$ up to $256$. 
Calculations for $T/Q\ge 0.035$ are also carried out for systems sufficiently large, up to $L=512$, to give results in the 
thermodynamic limit.

\begin{figure}
\centerline{\includegraphics[width=7cm, clip]{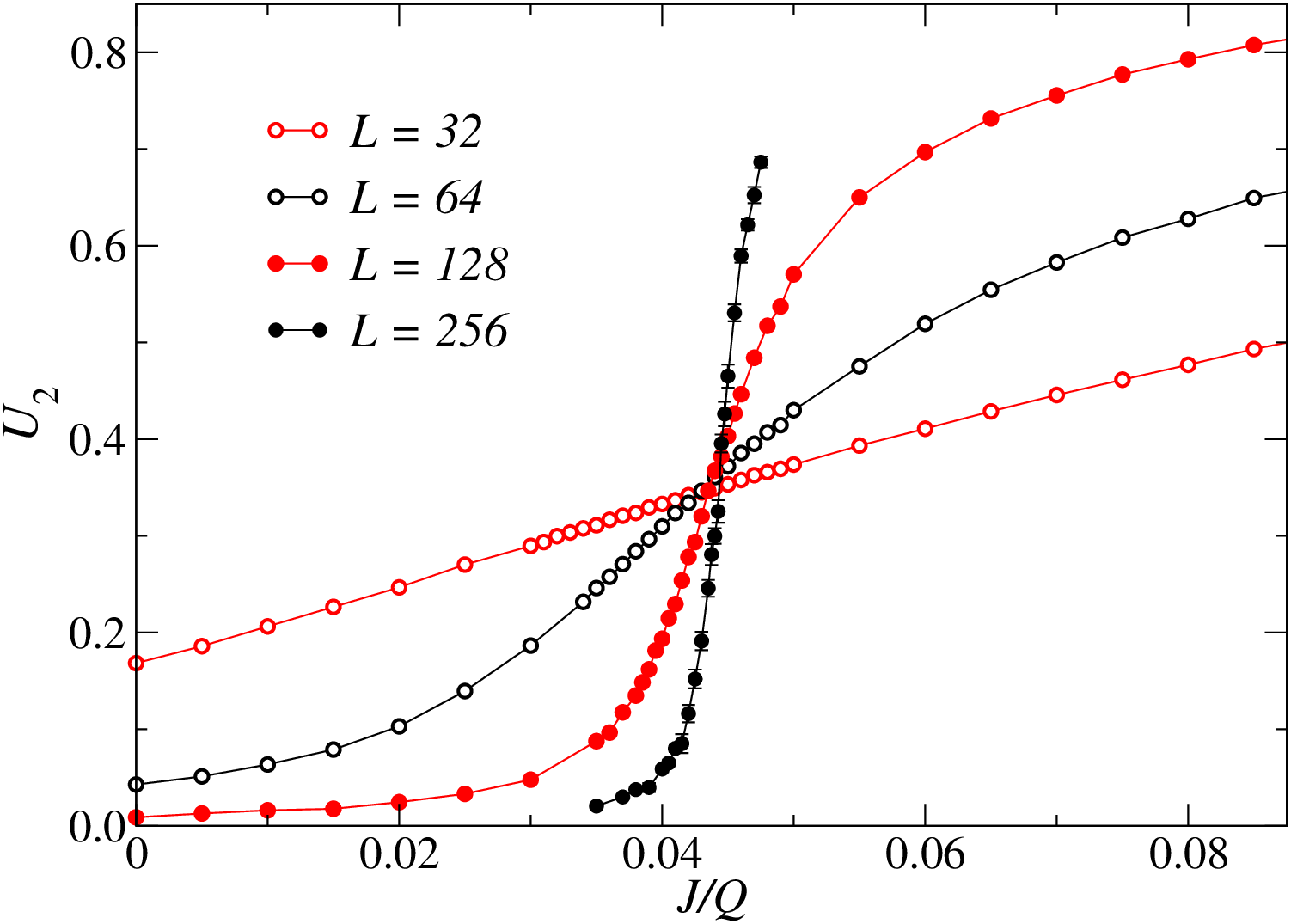}}
\vskip-2mm
\caption{(Color online) Binder cumulant for the sublattice magnetization as function of the coupling ratio for different
system sizes at inverse temperature $\beta=L$.}
\label{fig1}
\vskip-3mm
\end{figure}

The focus here will be on magnetic properties. The staggered magnetization $m_{s}$ is computed along the $z$ (quantization) axis. To extract 
the critical coupling ratio $(J/Q)_c$, and to address the issue of a possible first-order transition, consider first the Binder 
cumulant \cite{binder},
\begin{equation}
U_2 = \frac{5}{2} \left ( 1 - \frac{1}{3}\frac{\langle m_{sz}^4\rangle}{\langle m_{sz}^2\rangle^2} \right ),
\label{u2}
\end{equation}
which is defined so that $U_2 \to 0$ and $U_2 \to 1$ in an AF disordered and ordered state, respectively, when $L\to \infty$ (stemming from 
a Gaussian distribution of $|\vec m_s|$ around $|\vec m_s|=0$ and a $\delta$-function at $|\vec m_s|>0$, respectively). The factors 
in (\ref{u2}) correspond to $m_{sz}$ being one component of a three-dimensional vector $\vec m_s$. At a continuous transition, curves of 
$U_2$ versus $J/Q$ for different system sizes should intersect at the critical coupling, where normally $0 <U_2<1$ \cite{binder}. 
At a first-order transition, on the other hand, $U_2 \to -\infty$ when $L \to \infty$ \cite{binder}, following from a distribution 
with peaks at both $|\vec m_s|>0$ and $|\vec m_s|=0$ when the ordered and disordered phases coexist (with weight transferring rapidly 
between the peaks as the transition is crossed for large finite $L$). It should be noted that $U_2$ can be negative also at a continuous transition 
\cite{binder,beach}---only a divergence signals a first-order transition.

\begin{figure}
\centerline{\includegraphics[width=6.8cm, clip]{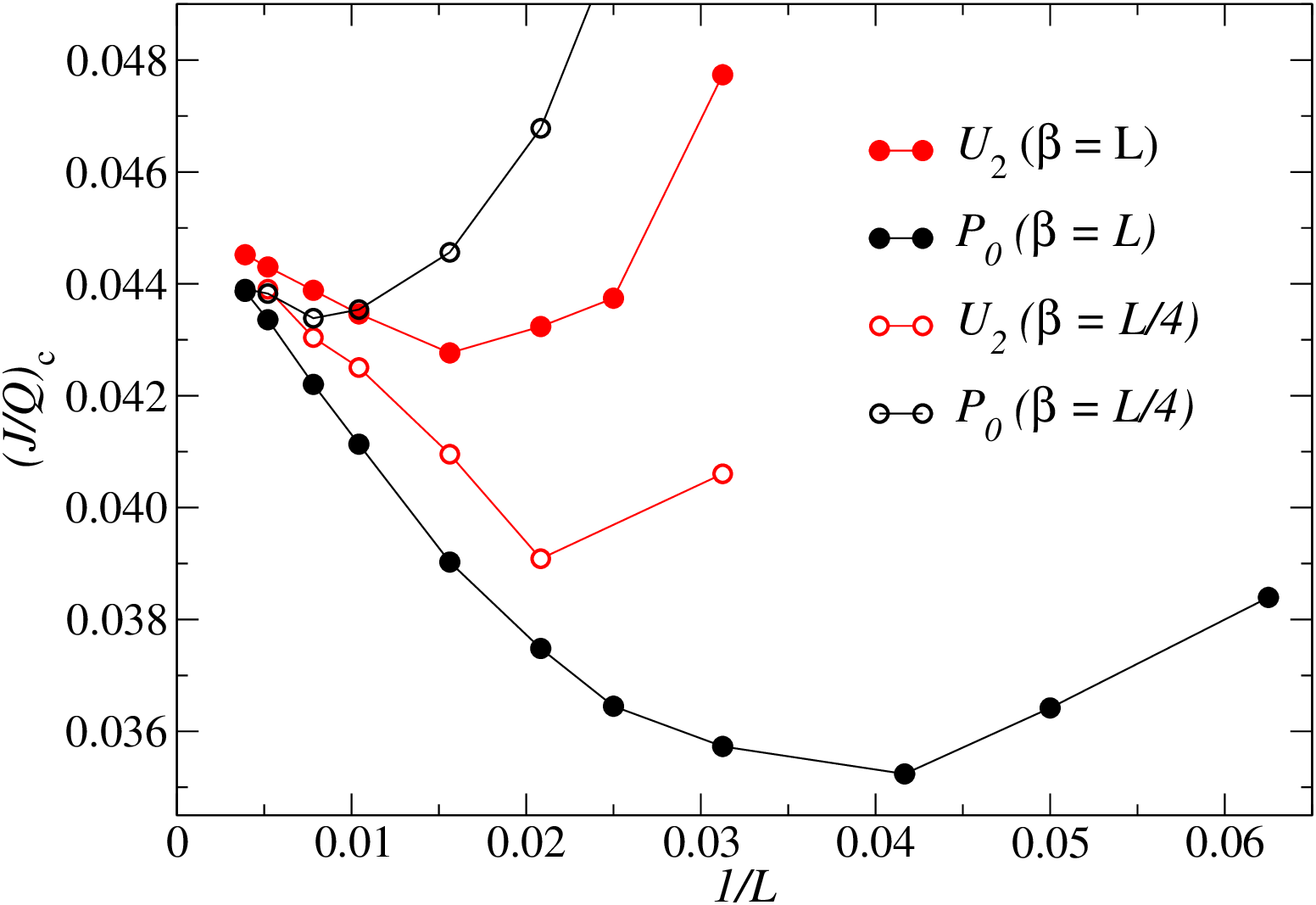}}
\vskip-2mm
\caption{(Color online) Critical couplings extracted from the crossing of $U_2(L)$ and $U_2(L/2)$ and from the winding number criterion 
$P_0=1/2$ in systems with $\beta=L$ and $L/4$.}
\label{fig2}
\vskip-3mm
\end{figure}

As seen in Fig.~\ref{fig1}, in the J-Q model there are no signs of $U_2$ becoming negative. The curves intersect at a point which
moves very slowly toward larger $J/Q$ with increasing system size. The critical coupling for $L\to \infty$ can be extracted by
extrapolating the crossing points for systems of size $L$ and $L/2$, as shown in Fig.~\ref{fig2}.

Fig.~\ref{fig2} also shows results for the size-dependent critical coupling suggested by Kuklov {\it et al.} \cite{kuklov1} and
used by Jiang {\it et al.} \cite{jiang}. It is based on the winding numbers,
\begin{equation}
W_{a} = \frac{1}{L}\sum_{p=1}^n J_{a}(p),
\end{equation}
where $J_a(p)$, $a=x,y$, is the spin current in lattice direction $a$ at location $p$ in an SSE configuration containing $n$ operators 
\cite{sandvik4}. In the case of the J-Q model, these currents take the values $J_a(p) \in \{ 0,\pm 1,\pm 2\}$. The ``temporal'' winding number 
is essentially the magnetization;
\begin{equation}
W_\tau = 2M_z,~~~~M_z=\sum_{i=1}^N S^z_i.
\end{equation}
The squared winding numbers are related to two important thermodynamic quantities; the spin stiffness,
\begin{equation}
\rho_s = \frac{1}{2\beta}\left (\left \langle W_x^2\right \rangle  + \left \langle W_y^2 \right \rangle \right ),
\end{equation}
and the uniform magnetic susceptibility,
\begin{equation}
\chi= \frac{\beta}{N} \left \langle M_z^2 \right \rangle = \frac{\beta}{4N}\left \langle W_\tau^2\right \rangle.
\end{equation}
For $L \to \infty$ and $T\to 0$, in a magnetically disordered (here VBS) phase $\rho_s \to 0$ and $\chi\to 0$, while in the AF phase
$\rho_s >0$ and $\chi > 0$. A possible  definition of the transition point (for finite $L$ and $\beta$) is the coupling at 
which the probability $P_0$ of all the winding numbers being zero is $1/2$ (or any fixed fraction) \cite{kuklov1}. Fig.~\ref{fig2} shows 
results obtained by interpolating $P_0$ for several $J/Q$ values. They extrapolate to the same $(J/Q)_c \approx 0.0445$ as the 
Binder cumulant crossings, but the corrections are larger. Note also that the size dependence varies significantly with the aspect ratio $\beta/L$. 
The results do not agree well with those of Jiang {\it et al.} \cite{jiang}, although $\rho_s$ and $\chi$ agree reasonably well for the system 
sizes available for comparisons. It is possible that $P_0$ is more sensitive to the Trotter approximation used in \cite{jiang}. Note also the 
non-monotonic size dependence in Fig.~\ref{fig2}. An $\ln(L)/L^3$ convergence of $(J/Q)_c$ was cited in \cite{jiang} as a sign of a first-order 
transition. The data fits were, however, based on only three system sizes. The behavior for larger lattices is clearly different.

\begin{figure}
\centerline{\includegraphics[width=7.5cm, clip]{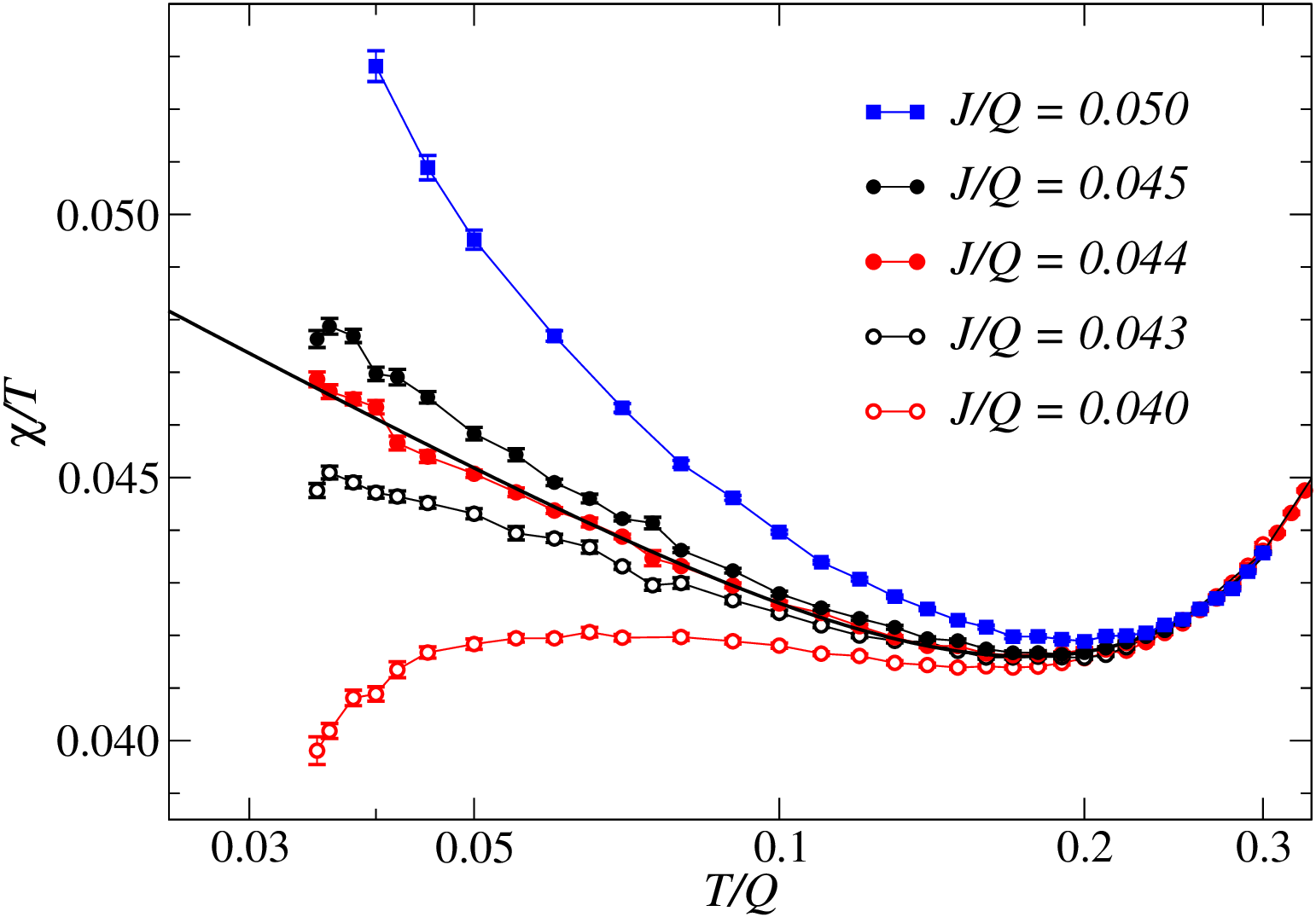}}
\vskip-2mm
\caption{(Color online) The uniform susceptibility divided by the temperature in the neighborhood of the critical point. 
The solid curve is of the form $\chi/T=a+b\ln(Q/T) + cT^2$.}
\label{fig3}
\vskip-3mm
\end{figure}

The $L \to \infty$ critical value $(J/Q)_c\approx 0.044$ is marginally higher than in previous studies. In particular, fitting the expected 
$z=1$ form $\chi \sim T$ of the susceptibility at $T>0$ ($L \to \infty$), Melko and Kaul found $(J/Q)_c \approx 0.038$. At higher $J/Q$ 
they found $\chi = a + bT$, as expected in the AF phase. Fig.~\ref{fig3} shows $\chi/T$ down to temperatures less than half of the lowest 
$T$ considered in \cite{melko}. At $J/Q=0.04$, while $\chi/T$ is roughly $T$-independent for $0.05 \alt T/Q \alt 0.2$, there is a drop at lower 
$T$, consistent with a spin-gapped phase. Close to the critical point there is no pure $\chi \propto T$ dependence at 
low $T$; instead the data exhibit a slow divergence, $\chi/T \approx a + b\ln(Q/T)$. The fanning-out of the data suggests that the logarithmic 
form is a critical separatrix between the expected $T\to 0$ behaviors in the VBS and AF phases.

\begin{figure}
\centerline{\includegraphics[width=6.25cm, clip]{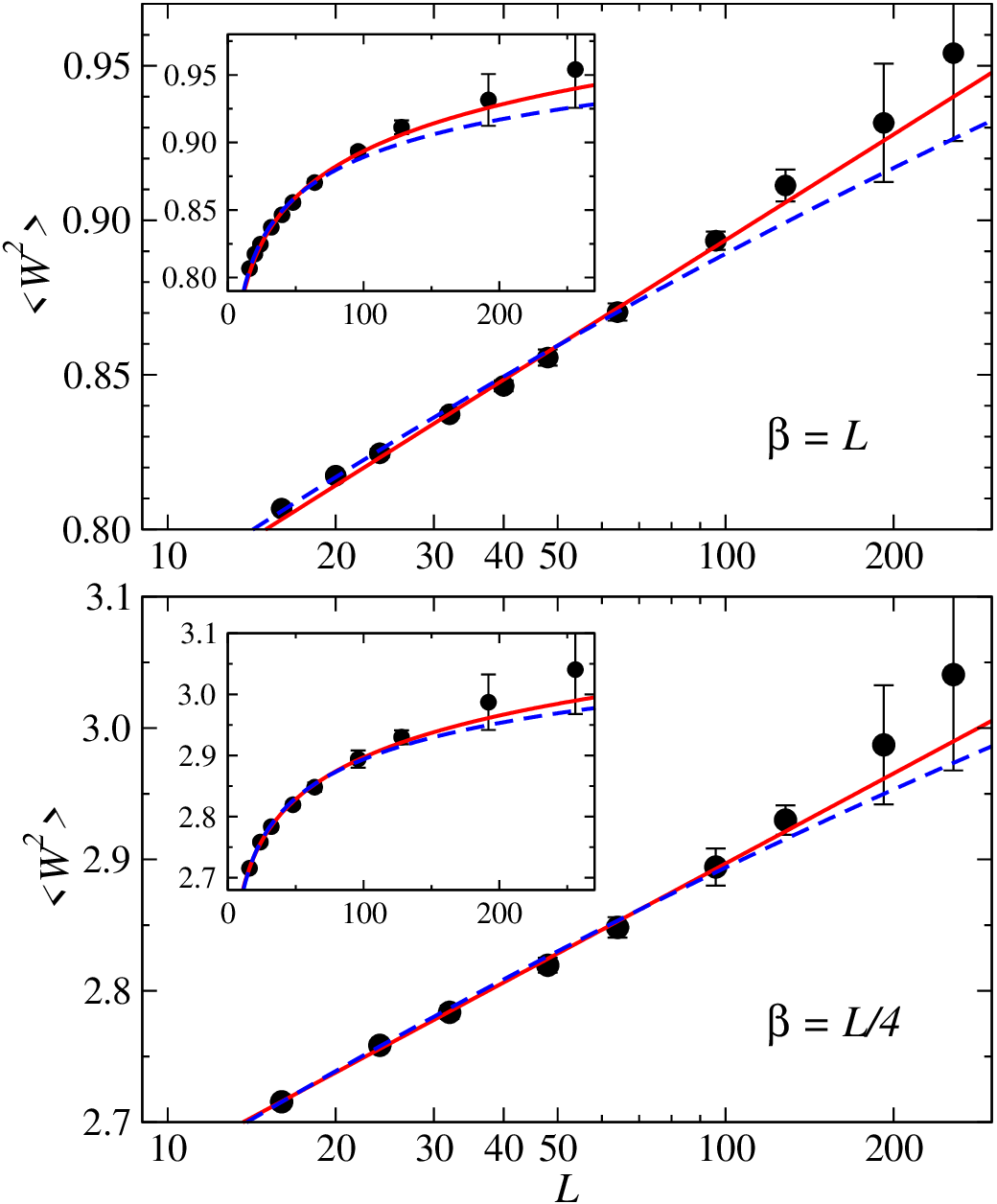}}
\vskip-2mm
\caption{(Color online) Size dependence of the total winding number at the $P_0(L)=1/2$ point for $\beta/L=1$ and $1/4$. The data are 
shown on log-lin (main panels) and lin-lin scales (insets). The solid and dashed curves are fits to forms $a + b\ln(L)$ and $c - dL^{-0.1}$, 
respectively.}
\label{fig4}
\vskip-3mm
\end{figure}

Another indication of logarithmic corrections comes from the total squared winding number,
\begin{equation}
\langle W^2\rangle = \langle W_x^2\rangle  + \langle W_y^2\rangle + \langle W_\tau^2\rangle = 2\beta \rho_s + \frac{4N}{\beta}\chi,
\label{w2def}
\end{equation}
for which Jiang {\it et al.} claimed an asymptotic linear divergence at the transition \cite{jiang}, as would be expected when AF and 
VBS phases coexist at a first-order transition. Fig.~\ref{fig4} shows the results of the present study. While $\langle W^2\rangle$ indeed
grows with $L$, it does so very slowly, consistent with a logarithmic divergence. There is no  plateau followed by a linear 
divergence---that conclusion \cite{jiang} seems to be based on an over-interpretation of noisy data. 

In principle, it is not possible to distinguish between a logarithm and a conventional scaling correction $\sim L^{-\omega}$ with a very small 
$\omega>0$. Fig.~\ref{fig4} shows fits with $\omega=0.1$ along with the logarithmic form---when $\omega \to 0$ the two forms coincide exactly. 
This comparison shows that if the corrections are conventional, then $\omega \alt 0.1$. This is true also for
the uniform susceptibility (Fig.~\ref{fig3}). 

\begin{figure}
\centerline{\includegraphics[width=7.5cm, clip]{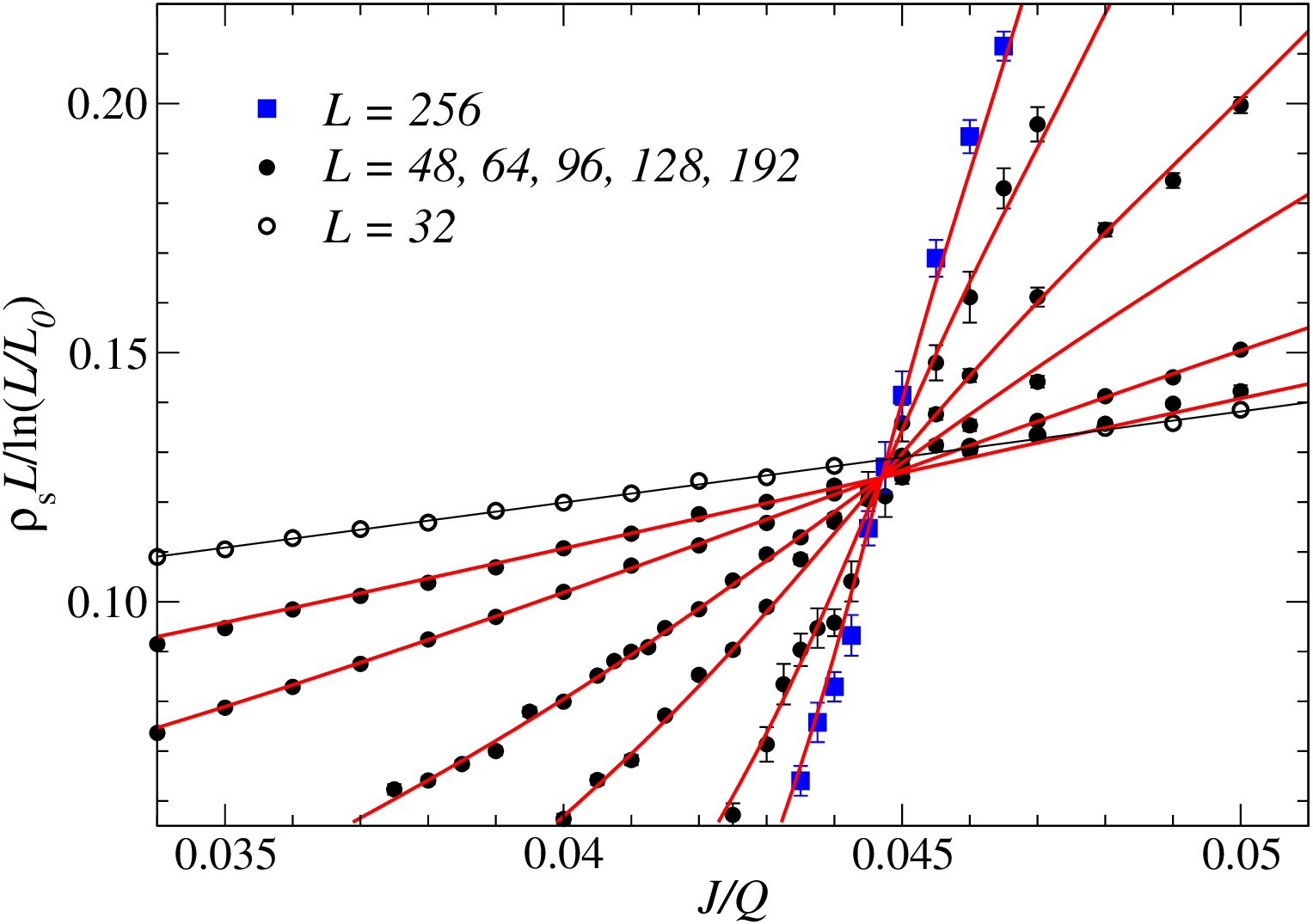}}
\vskip-2mm
\caption{(Color online) Scaling of the spin stiffness with a log-correction ($L_0=0.9$) for $\beta=L$ systems. The curves show a fit to a 
common polynomial $f[(J-J_c)L^{1/\nu}]$ with $J_c/Q=0.0447$ and $\nu=0.59$ (not including the $L=32$ data).}
\label{fig5}
\vskip-3mm
\end{figure}

Consider now the stiffness [not combined with $\chi$ as in (\ref{w2def})]. At a conventional $z=1$ critical point $\rho_s \sim 1/L$. In the present 
case the drift in crossing points of $\rho_SL$ curves for different $L$ is larger than what is normally \cite{wang} expected, but can be compensated 
by a logarithm, $\rho_s L/\ln (L/L_0)$, as shown in Fig.~\ref{fig5}. For $L \ge 48$ the curves intersect at a point, giving 
$(J/Q)_c = 0.0447 \pm 0.002$, in agreement with all the other results discussed above. Scaling fits away from the critical point give a correlation 
length exponent $\nu\approx 0.6$, but this is without considering possible corrections also to the conventional $L^{1/\nu}$ scaling. It is difficult 
to include logarithmic corrections in quantities where the leading exponent is not known, in contrast to $\rho_s$ and $\chi$ where $z=1$ governs
the leading behavior.

The conclusion of this study is that the AF--VBS transition in the J-Q model is continuous, but with significant corrections to the $z=1$ scaling that 
have not been discussed previously. The corrections appear to be logarithmic, although conventional scaling corrections $\sim L^{-\omega}$ with 
$\omega<0.1$ cannot be ruled out based on the numerical data alone. Regarding the possibility of a very weakly first-order transition, it should
be noted that rigorous proofs of continuous phase transitions are only available for a small number of exactly solvable models, yet accumulated 
numerical evidence of scaling (and experiments on natural systems), along with non-rigorous analytical calculations, have established a consensus 
that critical points are ubiquitous. The system volumes $\beta L^2$ used here for the J-Q model are similar to those in contemporary classical 
Monte Carlo simulations \cite{campostrini}. In the absence of any concrete signals of first-order behavior, the transition must therefore be 
regarded as continuous. 

The scaling corrections will hopefully stimulate further field-theoretical work to explain them. Scaling anomalies that could be logarithmic have 
been seen in Monte Carlo studies of the NCCP$^1$ action \cite{motrunich}, but it has also been claimed that this action always leads to a 
first-order transition \cite{kuklov2} (in which case a different field-theory for the J-Q model would have to be found). Marginal operators leading 
to logarithms appear in systems at their upper critical dimension, but this is not applicable here. Logarithmic corrections have been previously 
found in gauge field theories with fermions \cite{kim}. On the other hand, conventional power-law corrections due to irrelevant operators are always
expected, but here the subleading exponent $\omega$ would have to be very small, which has
not been anticipated (although the dangerously irrelevant operator causing the VBS has a small 
scaling dimension \cite{lou} and is a potential source of a small $\omega$). Studies of the SU($N$) 
generalization of the J-Q model would be useful to determine whether $N=2$ is a special case. QMC calculations have already been carried out for 
$N=3$ and $4$ \cite{lou}, but the quantities discussed here have not yet been investigated. 

A consequence of the findings presented here is that the anomalous VBS transition in U($1$) symmetric systems \cite{sandvik3} should be re-evaluated. 
Scaling deviations very similar to (but stronger than) those in the J-Q model were found, which in \cite{kuklov1,kuklov2} was interpreted as a 
first-order transition. Considering scaling corrections, this class of models as well may in the end have continuous transitions \cite{sandvik2}.

{\it Acknowledgments}---I would like to thank L. Balents, R. Kaul, V. Kotov, R. Melko, F. Nogueira, S. Sachdev, and T. Senthil for 
useful discussions. This work is supported by NSF grant No.~DMR-0803510. 

\null


\vskip-10mm
\begin{thebibliography}{00}

\bibitem{read}
N. Read and S. Sachdev, Phys. Rev. Lett. {\bf 62}, 1694  (1989).

\bibitem{senthil}
T. Senthil, A. Vishwanath, L. Balents, S. Sachdev, and M. P. A. Fisher, Science \textbf{303}, 1490 (2004).

\bibitem{motrunich}
O. I. Motrunich and A. Vishwanath, Phys. Rev. B {\bf 70}, 075104 (2004); arXiv:0805.1494.

\bibitem{sachdev}
S. Sachdev and X. Yin, Annals of Physics, {\bf 325}, 2 (2010).

\bibitem{sandvik1}
A. W. Sandvik, Phys. Rev. Lett. {\bf 98}, 227202 (2007).

\bibitem{sandvik2}
A. W. Sandvik, S. Daul, R. R. P. Singh, and D. J. Scalapino, Phys. Rev. Lett. {\bf 89}, 247201 (2002).

\bibitem{sandvik3}
A. W. Sandvik and R. G. Melko, Annals of Physics (N.Y.) {\bf 321}, 1651 (2006).

\bibitem{liang}
S. Liang, B. Doucot, and P. W. Anderson, Phys. Rev. Lett. {\bf 61}, 365 (1988).

\bibitem{signproblem}
P. Henelius and A. W. Sandvik, Phys. Rev. B {\bf 62}, 1102 (2000).

\bibitem{j1j2}
E. Dagotto and A. Moreo, Phys. Rev. Lett. \textbf{63}, 2148 (1989);
H. J. Schulz, T. Ziman, and D. Poilblanc, J. Phys. I \textbf{6}, 675 (1996).

\bibitem{sushkov}
O. P. Sushkov, J. Oitmaa, and Z. Weihong, Phys. Rev. B {\bf 63}, 104420 (2001).

\bibitem{evertz}
H. G. Evertz, Adv. Phys. {\bf 52}, 1 (2003).

\bibitem{white}
S. R. White, Phys. Rev. Lett. {\bf 69}, 2863 (1992).

\bibitem{analytical}
Analytical many-body techniques [V. N. Kotov, D. X. Yao, A. H. Castro Neto, and D. K. Campbell, Phys. Rev. B {\bf 80}, 174403 
(2009)] and approximate numerical approaches [L. Isaev, G. Ortiz, and J. Dukelsky, J. Phys. Cond. Mat. {\bf 22}, 016006 (2009)] have so far had 
only limited success with the J-Q model.

\bibitem{lou}
J. Lou, A. W. Sandvik, and N. Kawashima, Phys. Rev. B {\bf 80}, 180414(R) (2009).

\bibitem{melko}
R. G. Melko and R. K. Kaul, Phys. Rev. Lett. {\bf 100}, 017203 (2008);
R. K. Kaul and R. G. Melko, Phys. Rev. B {\bf 78}, 014417 (2008).

\bibitem{jiang}
F.-J. Jiang, M. Nyfeler, S. Chandrasekharan, and U.-J. Wiese, J. Stat. Mech., P02009 (2008).

\bibitem{kuklov1}
A. B. Kuklov, N. V. Prokof'ev, B. V. Svistunov and M. Troyer, Annals of Physics {\bf 321}, 1602 (2006).

\bibitem{kuklov2}
A. B. Kuklov, M. Matsumoto, N. V. Prokof'ev, B. V. Svistunov, and M. Troyer, Phys. Rev. Lett. {\bf 101}, 050405 (2008).

\bibitem{sandvik4}
A. W. Sandvik, Phys. Rev. B {\bf 59}, R14157 (1999).

\bibitem{ssenote}
In \cite{melko} a ``directed loop'' SSE method was used, whereas here a straight-forward generalization of ``deterministic loops'' 
\cite{sandvik4} are used.

\bibitem{nogueira}
F. S. Nogueira, Phys. Rev. B {\bf 77}, 195101 (2008).

\bibitem{metlitski}
M. A. Metlitski, M. Hermele, T. Senthil, and M. P. A. Fisher, Phys. Rev. B {\bf 78}, 214418 (2008).  

\bibitem{binder}
K. Binder, Phys. Rev. Lett. {\bf 47}, 693  (1981);
K. Binder and D. P. Landau, Phys. Rev. B {\bf 30}, 1477  (1984).

\bibitem{beach}
K. S. D. Beach, F. Alet, M. Mambrini, and S. Capponi, Phys. Rev. B {\bf 80}, 184401 (2009).

\bibitem{wang}
L. Wang, K. S. D. Beach, and A. W. Sandvik, Phys. Rev. B {\bf 73}, 014431 (2006).

\bibitem{campostrini}
M. Campostrini, M. Hasenbusch, A. Pelissetto, and E. Vicari, Phys. Rev. B {\bf 74}, 144506 (2006). 

\bibitem{kim}
D. H. Kim, P. A. Lee, and X.-G. Wen, Phys. Rev. Lett. {\bf 79}, 2109 (1997).

\end{thebibliography}
\end{document}